\documentclass[%
 aip,
 amsmath,amssymb,
 reprint,%
]{revtex4-1}

\usepackage{graphicx}% Include figure files
\usepackage{dcolumn}% Align table columns on decimal point
\usepackage{bm}% bold math
\usepackage[utf8]{inputenc}
\usepackage[T1]{fontenc}
\usepackage{mathptmx}
\usepackage{etoolbox}

\usepackage{xcolor}
\usepackage{hyperref}

\makeatletter
\def\@email#1#2{%
 \endgroup
 \patchcmd{\titleblock@produce}
  {\frontmatter@RRAPformat}
  {\frontmatter@RRAPformat{\produce@RRAP{*#1\href{mailto:#2}{#2}}}\frontmatter@RRAPformat}
  {}{}
}%
\makeatother
\begin{document}

\preprint{AIP/123-QED}

\title{A simple imaging solution for chip-scale laser cooling}
% Force line breaks with \\
\author{A. Bregazzi}
 \affiliation{SUPA and Department of Physics, University of Strathclyde, G4 0NG, United Kingdom}

\author{P. F. Griffin}%
\affiliation{SUPA and Department of Physics, University of Strathclyde, G4 0NG, United Kingdom}

\author{A. S. Arnold}
\affiliation{SUPA and Department of Physics, University of Strathclyde, G4 0NG, United Kingdom}

\author{D. P. Burt}
 \affiliation{Kelvin Nanotechnology, University of Glasgow, G12 8LS, United Kingdom}

\author{G. Martinez}
\affiliation{University of Colorado, Department of Physics, Boulder, Colorado 80309, USA}
\affiliation{National Institute of Standards and Technology, Boulder, Colorado 80305, USA}

\author{R. Boudot}
\affiliation{FEMTO-ST, CNRS, 26 Chemin de l‘Epitaphe, 25030 Besan\c con, France}

\author{J. Kitching}
\affiliation{National Institute of Standards and Technology, Boulder, Colorado 80305, USA}

\author{E. Riis}
\affiliation{SUPA and Department of Physics, University of Strathclyde, G4 0NG, United Kingdom}

\author{J.~P.~McGilligan}
\affiliation{SUPA and Department of Physics, University of Strathclyde, G4 0NG, United Kingdom}
\affiliation{Kelvin Nanotechnology, University of Glasgow, G12 8LS, United Kingdom}
 \email{james.mcgilligan@strath.ac.uk}

\date{\today}% It is always \today, today,
             %  but any date may be explicitly specified

\begin{abstract}
%The miniaturisation of laser cooling platforms to the chip scale has been largely aided by the coupling of micro-fabricated vacuum cells and the grating magneto-optical trap (GMOT). Further volume reductions of this apparatus have been frustrated by the complexity required to image the trapped atoms while overcoming the reduced optical access and surface scatter in close proximity to the cold atom ensemble. We demonstrate a simple stacked scheme that enables absorption imaging through a hole in the grating surface, permitting both trapping and imaging of the atoms using a single incident laser beam. The through-hole imaging is used to characterise the impact of a 3~mm thick micro-fabricated cell on the optical overlap volume of the GMOT, with an outlook to an optimised atom number in low volume systems.

We demonstrate a simple stacked scheme that enables absorption imaging through a hole in the surface of a grating magneto-optical trap (GMOT) chip, placed immediately below a micro-fabricated vacuum cell. The imaging scheme is capable of overcoming the reduced optical access and surface scatter that is associated with this chip-scale platform, while further permitting both trapping and imaging of the atoms from a single incident laser beam. The through-hole imaging is used to characterise the impact of the reduced optical overlap volume of the GMOT in the chip-scale cell, with an outlook to an optimised atom number in low volume systems.

\end{abstract}

\maketitle

Micro-fabricated physics packages based on the measurement of thermal atomic ensembles have been integrated in metrological instruments, ranging from magnetometers to interferometers and clocks \cite{Knappe2004,Kitching2018,Carle2021_microfab_clock}. Most such instruments rely on spin transitions in atoms, and require the presence of buffer gas mixtures or cell wall coatings to reduce the relaxation rate due to wall collisions. For clocks, the presence of these buffer gases causes temperature dependent frequency offsets and require careful temperature stabilisation to achieve good medium- and long-term stability \cite{Kozlova2011_bufer_gas,Hasegawa2011_buffer_gas,Seltzer2009_cell_coating}. The use of cold atom ensembles avoids these difficulties and can result in an increased interaction time and improved absolute accuracy compared to their thermal atom counterparts.

While the transition to cold atoms offers clear advantages over thermal atom packages \cite{Eckel_2018}, the additional experimental size and complexity associated with laser cooling has limited the deployability and application range of cold atom devices \cite{Rushton2014,marlow2021review,MuClock}. Significant efforts have been made on the miniaturisation of cold atom components to facilitate portability for next-generation atomic sensors \cite{burrow2021,little2021,ball_lens_MOT,Nshii2013,Ravenhall21,Kang:19,McGehee_2021, eckelsr}. However, the fabrication complexity of many of these components remains unfavourable for mass production, preventing their adoption in commercial applications.  

%Significant efforts have been made on the miniaturisation of cold atom components to facilitate portability for next-generation atomic sensors \cite{burrow2021,little2021,ball_lens_MOT}. However, the mass production and fabrication complexity remain unfavourable for commercial applications. Recent work has demonstrated a dramatic reduction of the cooling platform to the chip-scale by combining a silicon-glass micro-machined vapour cell with a diffractive optical element that redirects a single incident beam into the components required for laser cooling \cite{McGilligan2020}. 
Recent work has demonstrated a dramatic reduction of the cooling platform to the chip-scale by combining an anodically bonded glass-silicon-glass vapour cell with a diffractive optical element that redirects a single incident beam into the components required for laser cooling \cite{McGilligan2020}. However, such systems have demonstrated that the detection of cold atoms is made difficult by the reduced optical access of the cell and light scattering from the grating and cell surfaces. As such, the authors required adopting a non-trivial two-photon spectroscopy scheme for improved detection, greatly increasing both the size and complexity of the optical system \cite{McGilligan2020,scholten}.

In this letter we demonstrate a simple imaging solution for chip-scale laser cooling platforms. The stackable structure of the apparatus provides simplicity in alignment, as well as enabling future scalability for device mass production. A hole is laser cut in the centre of the grating chip to enable on-axis absorption imaging from the cooling beam without significantly degrading atom trapping. Although grating chips with a central hole have been used to reduce surface reflections \cite{imhof}, as a source for cold electrons \cite{jimion}, and for Zeeman slowing with an alkali \cite{barkerNIST} and alkaline-earth metal source \cite{srgmot}, the impact of the reduced grating surface area on the MOT number has not yet been quantified or used as an imaging axis. The ability to measure the trapped atom number in such a chip-scale system is used to characterise the impact of the 3~mm internal height constraint on the optical overlap volume, with an outlook to an optimised atom number.

\begin{figure*}[t]
\includegraphics[width=1\textwidth]{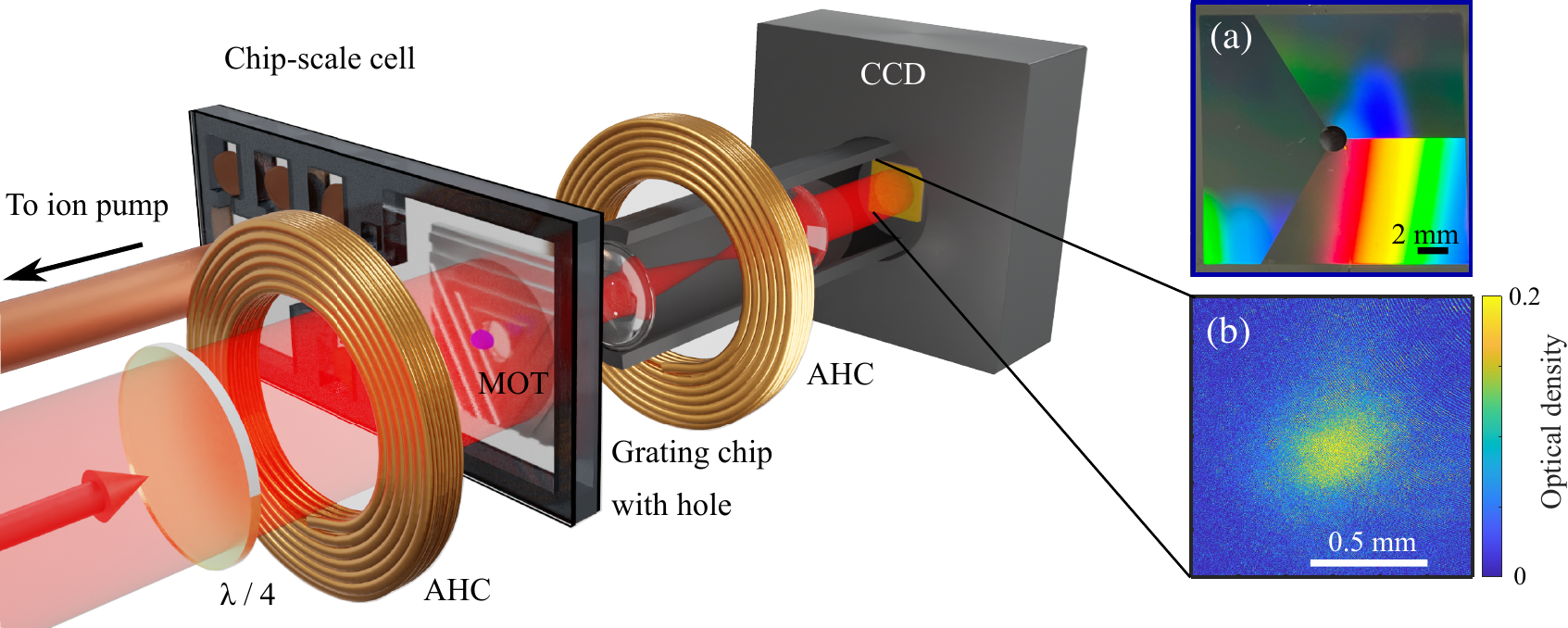}% Here is how to import EPS art
\caption{\label{fig:exp_setup_and_MOT_image} Experimental set-up. CCD: charge coupled device, AHC: anti-Helmholtz coils, $\lambda/4$: quarter-wave plate. (a): Image of the grating chip with a central hole removed. (b): An on-axis, through-hole absorption image of $\approx$\,2$\times$10$^5$ atoms within the chip-scale platform. This image was taken with a peak beam intensity of 115~$\mu$W/cm$^2$ at zero detuning.}
\end{figure*}

A simplified schematic for the cooling and imaging of $^{87}$Rb is shown in Fig~\ref{fig:exp_setup_and_MOT_image}. The incident light is derived from a single volume-Bragg-grating laser (VBG) frequency stabilised using saturated absorption spectroscopy. A double-pass acousto-optic modulator (AOM) shifts the light frequency to be approximately 8~MHz red detuned from the $780$~nm $D_2$ $F=2\rightarrow F'=3$ cycling transition, while also enabling frequency and intensity control for the imaging process. Re-pumping is achieved by modulating a free-space electro-optic modulator (EOM) at 6.5~GHz to generate 5~$\%$ sidebands on the carrier frequency. The light is then fibre coupled into a single-mode, polarization maintaining fibre and passed to the cooling platform. 

From the fibre, we expand an $\approx$30~mW beam to a $1/e^2$ radius of $\approx1.6$~cm to flatten the intensity distribution at the grating surface. The trap beam is then circularly polarised with a quarter wave-plate and aligned onto the grating chip, mounted externally to the actively pumped chip-scale cell. A pair of anti-Helmholtz coils are used to produce a trapping field within the cell volume, with an axial gradient of $\approx$1.5~mT/cm (15~G/cm).

The cell, shown in Fig.~\ref{fig:exp_setup_and_MOT_image}, is composed of a $3$~mm thick silicon wafer, sandwiched between two anodically bonded aluminosilicate glass wafers of 0.7~mm thickness \cite{dellis2016lowHe}. The cell was fabricated with a square central region of dimensions 2.5~cm$\times$2.5~cm, to enable cooling with a 2~cm wide grating chip. The upper window of the cell is mechanically drilled with a 5~mm hole to enable active pumping from an ion pump through a copper pinch-off tube, adhered to the upper cell surface. Square cavities are cut in the silicon walls to house non-evaporable getters (NEGs) for future passive pumping measurements.

The grating chip, shown in Fig.~\ref{fig:exp_setup_and_MOT_image} (a), has a 1100~nm period in a tri-segment geometry. The central region of the chip is laser cut to provide a through-hole axis for absorption imaging. This central region of the grating plays a limited role in forming the trap overlap volume while also providing a unfavourable zeroth order reflection, such that its removal has a minimal impact on atom number. A two-lens imaging telescope and camera are placed behind the grating hole to detect atomic absorption from the GMOT. 

While extracting atom number with the through-hole system, fluorescence measurements from the MOT loading curves were simultaneously measured using a separate detection telescope with a spatially selective focal plane (described in Ref.~\cite{BoudotPassive}) at an angle of around 30\textdegree~to the grating surface (not shown in Fig.~\ref{fig:exp_setup_and_MOT_image}). The loading curves were used for the extraction of background pressure at the MOT location using the relation between MOT lifetime and background rubidium pressure \cite{arpornthip2012pressuremeasurment, moore2015pressuremeasurement, burrow2021}, while also checking the validity of the through-hole measured atom number. We note that the atom number measured from fluorescence imaging at this angle well matched the atom number extracted from conventional orthogonal fluorescence imaging. Additionally, an average ratio of 1.6$\pm$0.4 was observed between the calculated atom number from the through-hole absorption and orthogonal fluorescence during our data sets. However, the alignment of the angled fluorescence imaging axis was complicated due to the surface scatter and diffracted orders further restricting the available imaging angles and contributing to an increased noise level on the atom number extraction. The complexity in aligning the camera at a position and angle that minimises the impact of surface scatter does not meet the needs of a simple, mass-producible device.  

The experimental procedure is initiated by turning the trap coils on, followed by a 500~ms loading time at $8$~mW/cm$^2$ peak intensity. The 500~ms load time enables resolving the vacuum pressure down to 1.4~$\times$10$^{-6}$~Pa (1.4$\times$10$^{-8}$~mbar) \cite{McGilligan2017}. The trap coils are then turned off while concurrently decreasing the incident beam below saturation to around 115~$\mu$W/cm$^2$ and bringing the frequency on resonance. This serves the dual purpose of reducing image distortion due to diffraction effects within the atom cloud while also maximising the signal contrast. We note that the measured atom number from this method did not differ with the addition of a static magnetic field along the imaging axis. An initial image $I_1$ is taken with the MOT present for an exposure time of 25~$\mu$s. We then wait 100~ms such that the trapped atoms are no longer present before taking a second image $I_2$. Finally the trap beam is turned off with an extinction ratio of 62~dB and a dark background image $I_3$ is acquired. These three images are then processed in terms of optical density (OD) using the equation $\rm{OD}=\rm{ln}\Big(\frac{\textit{I}_{1}-\textit{I}_{3}}{\textit{I}_{2}-\textit{I}_{3}}\Big)$. An objective lens of focal length $f_{1}=15$~mm and an imaging lens $f_{2}=30$~mm, arranged in $2f_1:2f_2$ configuration, is used to image the atoms, with the objective placed directly behind the hole laser cut in the centre of the grating. This arrangement of lenses provides an improved signal-to-noise ratio for data extraction \cite{smithabsorption}. 

In the present work, the stretched state saturation intensity $I_{\rm{sat}}$=1.67~mW/cm$^2$ was used for calculating the trapped atom number from the fluorescence and absorption methods, as is consistent with previous publications of our group \cite{Nshii2013}. Additionally, we found good agreement between the calculated atom numbers from the two imaging methods with the stretched state saturation intensity, compared to the average of all polarizations and magnetic sub-levels, $I_{\rm{sat}}$=3.57~mW/cm$^2$. A saturation parameter, $S=\frac{I}{I_{\rm{sat}}}$, sums over the intensities of the single input trap beam and the three diffracted beams from the grating to account for the total intensity the atoms experience. This inclusion modifies $S$ such that $S=(1+n\eta_1\sec\theta)\frac{I_{\rm{im}}}{I_{\rm{sat}}}\approx\frac{2.4I_{\rm{im}}}{I_{\rm{sat}}}$, where $I_{\rm{im}}$ is the imaging light intensity, $n$ is the number of diffracted first orders interacting with the atoms, $\eta_1$ is the efficiency of the first diffracted order, and $\theta$ is the angle of diffraction. An example of the obtained MOT image is shown in Fig.~\ref{fig:exp_setup_and_MOT_image} (b), where 2$\times10^{5}$ atoms are trapped in the chip-scale cell.

\begin{figure}[t]
\includegraphics[width=0.48\textwidth]{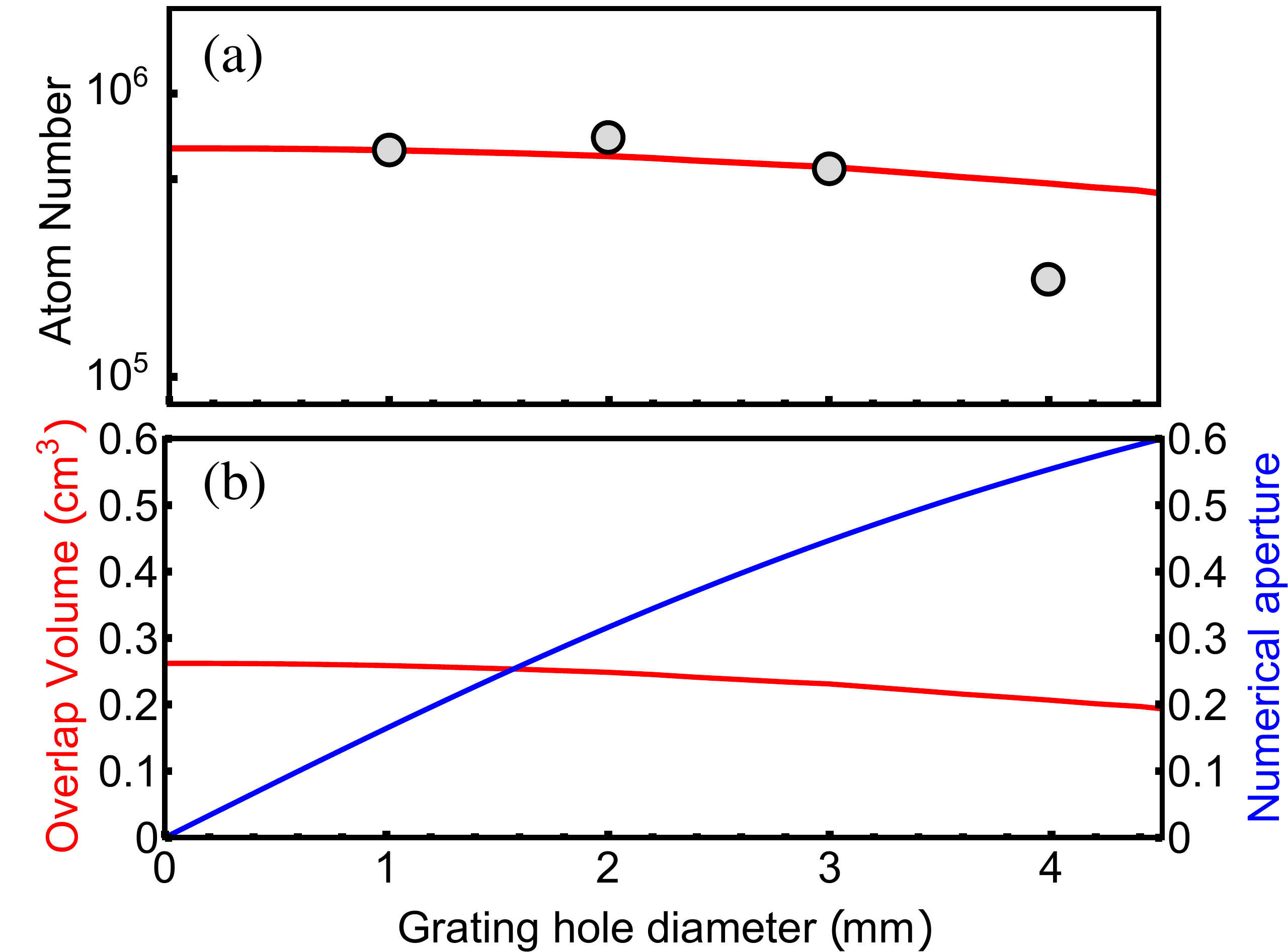}% Here is how to import EPS art
\caption{\label{fig:gratinghole} Critical parameters as a function of the grating centre hole diameter for an incident 2~cm beam diameter. (a): The measured atom number from orthogonal fluorescence imaging in the conventional glass cell. The measured error bars are smaller than the shown data points. The red curve shows the normalized $N\propto V^{1.2}$ for the overlap volume shown in (b). (b): Theoretically calculated optical overlap volume and numerical aperture from the grating chip assuming that the grating hole is the aperture stop of the imaging system.}
\end{figure}

Our investigation of the impact of removing the central region of the grating chip was initially carried out in a conventional glass vacuum chamber ($2.5$~cm$\times2.5$~cm$\times10$~cm with 1~mm thick glass walls) using fluorescence imaging, with the results shown in Fig.~\ref{fig:gratinghole}. We fabricated 5 identical grating chips, and laser cut the central regions for a hole size ranging from 1~mm to 5~mm. For each subsequent measurement of the atom number, the grating chip was carefully implemented immediately below the same vacuum cell, under the same conditions as the previous grating chip. We found that a hole diameter up to 3~mm did not significantly degrade the measured atom number, shown in Fig.~\ref{fig:gratinghole} (a), due to the minimal impact that such a hole size has on the total optical overlap volume, shown in Fig.~\ref{fig:gratinghole} (b). The theoretical optical overlap volume is numerically integrated for a 2~cm beam diameter, incident upon a 1100~nm period tri-segmented grating chip with a central hole of varying diameter. As the grating hole was increased to 4~mm, the atom number decreases, reducing to the point that no MOT was detected for a 5~mm hole diameter, even with additional alignment of the MOT parameters. The atom number drops faster than would be expected from the decrease in the overlap volume, likely due to the additional sensitivity of the radiation pressure balance from the grating chip. This is emphasized by the red curve shown in Fig.~\ref{fig:gratinghole} (a), where the theoretically calculated overlap volume is scaled to the atom number through $N\propto V^{1.2}$ and normalised to the measured atom number for the 1~mm hole. For a grating hole diameter of 4~mm, a factor of $\sim$2 difference between the expected and measured atom number is observed. However, the increased hole diameter aids an improved numerical aperture, also plotted in Fig.~\ref{fig:gratinghole} (b) for a MOT position of 2~mm above the grating surface, such that a larger hole diameter is favourable for imaging with the small atom numbers that have been observed in passively pumped vacuum cells \cite{BoudotPassive}. Taking this into account, a 3~mm hole diameter was selected for through-hole imaging.

Once the through-hole imaging was well aligned and in focus, the glass cell was replaced by the actively pumped chip-scale vacuum cell. The chip-scale cell was pumped down to an initial pressure of 10$^{-6}$~Pa (10$^{-8}$~mbar), measured using an ion pump. The validity of the ion pump reading was later verified with the pressure calculated from MOT loading curves. Following pump down, a resistively heated dispenser within the larger vacuum chamber was used to provide a moderate rubidium density. With the grating overlap volume aligned within the chip-scale cell, an image of the cold atoms was extracted from the through-hole scheme, highlighted in Fig.~\ref{fig:exp_setup_and_MOT_image}. The initial atom number measurement yielded a total of 2$\times10^{5}$ atoms, which was an order of magnitude lower than what had been observed for the glass cell.

\begin{figure}[t]
\includegraphics[width=0.45\textwidth]{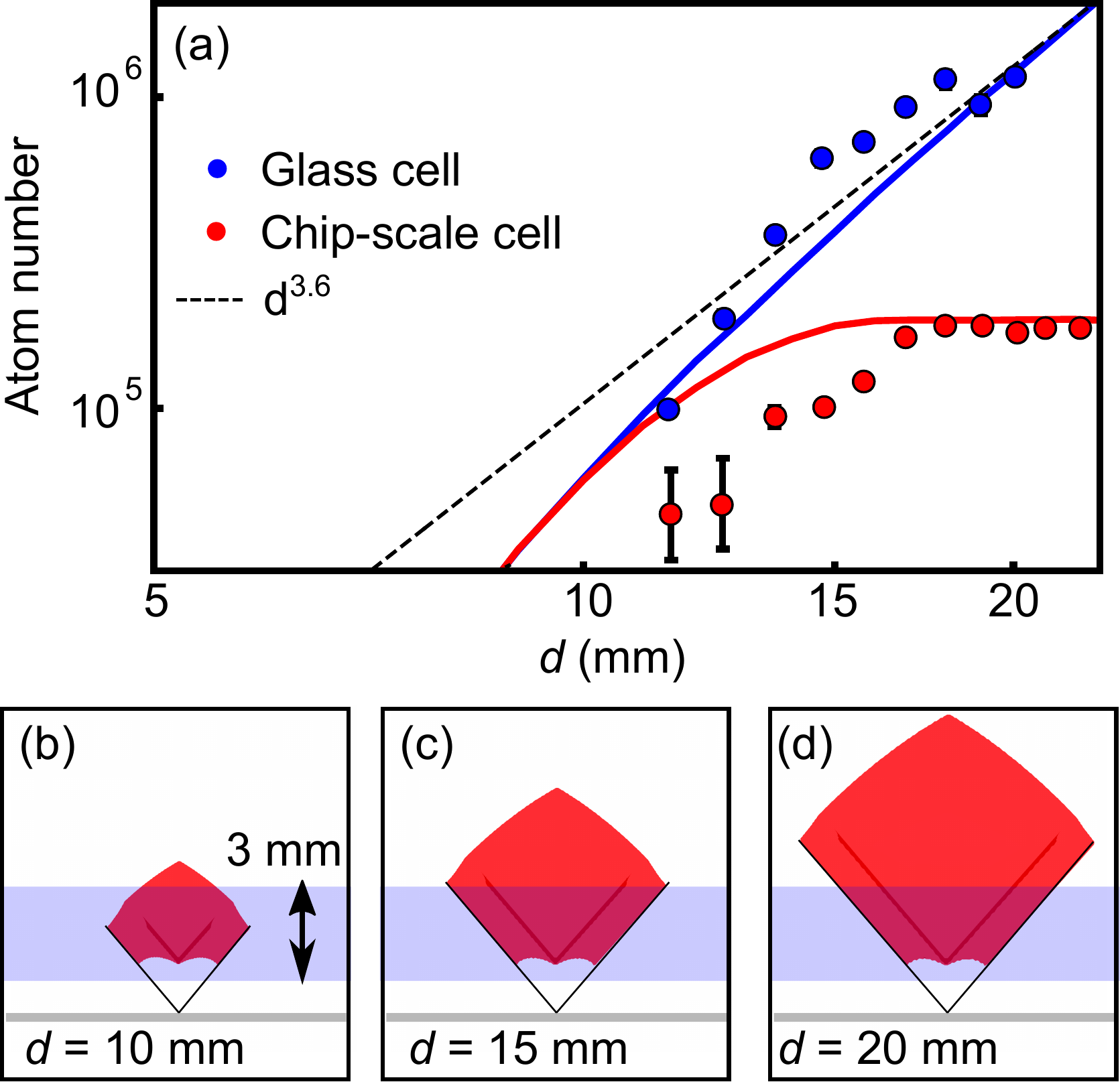}
\caption{\label{fig:N_vs_beam_diameter}(a): Atom number measured from absorption imaging as a function of the incident beam diameter, $d$. Blue/red data points represent measurements made in a glass/chip-scale cell. Error bars were calculated from 5 subsequent atom number measurements at each beam diameter. Solid curves are theoretical expectations of the atom number, determined from numerical integration of the optical overlap volume from the grating chip as a function of $d$. The normalised theoretical atom number is pinned to the maximum measured atom number for the glass cell. The red line shows an overlap volume that is restricted to a 3~mm height due to the vacuum volume of the chip-scale cell. The dashed line represents a $d^{3.6}$ fit to the atom number in the glass cell. (b)-(d) illustrate the optical overlap volume of the grating chip as a function of $d$ with a highlighted 3~mm height of the chip-scale cell. The grey line highlights the grating chip position.}
\end{figure}

To investigate the cause of the reduced atom number, the diameter of the incident cooling light onto the grating surface, $d$, was reduced while measuring the trapped atom number for both the glass and chip-scale cells. This measurement provides an insight to the impact of the reduced vacuum volume of the chip-scale cell on the grating chip's optical overlap volume, and therefore trapped atom number. During this process a comparable background vapour density of $\approx3.1\times10^6$~cm$^{-3}$ and total pressure of $\approx3\times10^{-5}$~Pa ($\approx3\times10^{-7}$~mbar) was maintained in both vacuum systems. The resulting data set is shown in Fig.~\ref{fig:N_vs_beam_diameter} (a) in blue and red for the glass and chip-scale cells, respectively.

At the largest values of $d$ an order of magnitude difference in the atom number is observed between the glass and chip-scale cell, with the glass cell data reaching a maximum around $10^6$ atoms, in line with previous atom numbers measured with a similar incident intensity \cite{mcgilligan15}. Concurrently, the red data set extracted from the chip-scale cell reaches a maximum around 10$^5$ atoms. When $d$ is reduced from this maximum atom number, there is initially no impact on the optical overlap volume, since the majority of the overlap volume is outside the cell's internal thickness and therefore cannot play a role in the cooling process. As $d$ is decreased below $\approx17$~mm, the overlap volume is reduced below the internal height of the chip-scale cell, contributing to a degradation of the atom number.

The impact of the reduced optical overlap volume is emphasised by the theoretical atom number expected from the grating MOT as a function of the incident beam diameter, shown with solid curves in Fig.~\ref{fig:N_vs_beam_diameter} (a). This atom number, $N$, is derived from a numerical integration of the optical overlap volume, $V$, for a $1100$~nm period linear grating, where $N\,\propto\,V^{1.2}\,\propto\,d^{3.6}$\,\cite{Nshii2013}. The red curve represents the overlap volume from the grating chip with a 3~mm diameter hole at its centre. A vertical restriction is placed on the integration to account for the 3~mm thickness of the silicon frame. The overlap volume is then converted to a normalised atom number with a maximum corresponding to the largest measured atom number from the glass cell. A total potential overlap volume of $0.23$~cm$^3$ is possible for a 2~cm diameter incident upon a grating chip with a 3~mm diameter hole, held 1~mm from the cell window. However, when the grating is mounted similarly for the chip-scale cell, only 21$\%$ of the relative overlap volume coincides within the cell vacuum volume. This process is illustrated in Fig.~\ref{fig:N_vs_beam_diameter} (b)-(d), where the numerically calculated geometry of the optical overlap volume is shown for different values of $d$. The region of the overlap volume occupied by the 3~mm tall chip-scale cell is highlighted to show the clipping of the overlap volume as a function of $d$.

The dashed black line in Fig.~\ref{fig:N_vs_beam_diameter} shows a $d^{3.6}$ scaling of the atom number, expected for a constant intensity of the incident beam \cite{wieman1992,Gibble:92} that expands in three-dimensions. Previous studies have demonstrated that significantly smaller overlap volumes for 6-beam MOTs \cite{Hoth13} and pyramid MOTs \cite{Pollock_2011} are reduced to a $d^6$ scaling, as the atom number is then limited by the stopping distance of the overlap volume rather than the cooling light intensity that dominates the process for larger beam diameters. We note that the measured atom number in the chip-scale cell is lower than the theoretical $V^{1.2}$, shown as a solid red curve. The cause of this is currently unknown and will be investigated in future studies.  

When limited by atom shot noise, the stability of cold atom sensors scales inversely with the square-root of the atom number. In this case, an order of magnitude improvement in the chip-scale cell would be favourable. Expanding the overlap volume simulation to a 5~mm thick silicon frame enables a $71\%$ overlap between the optical overlap volume and the vacuum volume of the chip-scale cell when the grating is held 1~mm from the cell lower window. Using identical scaling parameters to the 3~mm chip-scale cell curve shown in Fig.~\ref{fig:N_vs_beam_diameter} (a) results in an achievable atom number of 10$^6$ atoms for a 2~cm incident beam diameter in this thicker cell geometry. Thus, we plan to fabricate thicker silicon cells in the near-future to enable a wider range of applications for this chip-scale laser cooling platform. Additionally, a steeper angle of diffraction with a two-dimensional grating geometry would provide an overlap volume that is largest at the chip surface for an improved atom number within the reduced vacuum volume compared to the tri-segmented chip used here. The improved atom number will enable studies into passive pumping with NEGs in an aluminosilicate based chip-scale cell \cite{BoudotPassive}.

In conclusion, we have demonstrated a simple imaging solution for atom number extraction in a chip-scale cooling platform. The laser cut hole in the grating chip has shown a negligible atom number degradation, while enabling a sufficient numerical aperture for imaging small atom numbers in future experiments. We have shown that the 3~mm height of the micro-fabricated cell reduces the achievable overlap volume, and hence atom number, when coupled with a micro-fabricated grating chip. However, moving to the machining of thicker silicon wafers will enable a work around to bring this platform to a competitive stance for quantum technologies. In addition it may also be advantageous to apply an anti-reflection coating to the cell windows during fabrication due to the increased etaloning observed in the through-hole images. While the effect of these fringes on the images was minimal, especially for larger MOTs, and can be mitigated through alignment there are also several fringe removal algorithms that can be applied in post-processing to further reduce their impact \cite{fringe_removal_Niu2018,fringe_removal_Xiong2020}.

\begin{acknowledgments}
The authors acknowledge funding from Defence Security and Technology Laboratory, Engineering and Physical Sciences Research Council (EP/T001046/1), and Defence and Security Accelerator. A. B. was supported by a PhD studentship from the Defence Science and Technology Laboratory (Dstl). The authors gratefully acknowledge W. R. McGehee and D. S. Barker (NIST) for the careful reading of the manuscript before submission. Additionally, the authors would like to thank B. Lewis and M. Himsworth for useful conversations. 
\end{acknowledgments}

\section*{Data Availability Statement}
The data that support the findings of this study are available from the corresponding author upon reasonable request.

%\nocite{*}
\bibliography{Ref}% Produces the bibliography via BibTeX.

%merlin.mbs aipnum4-1.bst 2010-07-25 4.21a (PWD, AO, DPC) hacked
%Control: key (0)
%Control: author (8) initials jnrlst
%Control: editor formatted (1) identically to author
%Control: production of article title (0) allowed
%Control: page (1) range
%Control: year (1) truncated
%Control: production of eprint (0) enabled
\begin{thebibliography}{37}%
\makeatletter
\providecommand \@ifxundefined [1]{%
 \@ifx{#1\undefined}
}%
\providecommand \@ifnum [1]{%
 \ifnum #1\expandafter \@firstoftwo
 \else \expandafter \@secondoftwo
 \fi
}%
\providecommand \@ifx [1]{%
 \ifx #1\expandafter \@firstoftwo
 \else \expandafter \@secondoftwo
 \fi
}%
\providecommand \natexlab [1]{#1}%
\providecommand \enquote  [1]{``#1''}%
\providecommand \bibnamefont  [1]{#1}%
\providecommand \bibfnamefont [1]{#1}%
\providecommand \citenamefont [1]{#1}%
\providecommand \href@noop [0]{\@secondoftwo}%
\providecommand \href [0]{\begingroup \@sanitize@url \@href}%
\providecommand \@href[1]{\@@startlink{#1}\@@href}%
\providecommand \@@href[1]{\endgroup#1\@@endlink}%
\providecommand \@sanitize@url [0]{\catcode `\\12\catcode `\$12\catcode
  `\&12\catcode `\#12\catcode `\^12\catcode `\_12\catcode `\%12\relax}%
\providecommand \@@startlink[1]{}%
\providecommand \@@endlink[0]{}%
\providecommand \url  [0]{\begingroup\@sanitize@url \@url }%
\providecommand \@url [1]{\endgroup\@href {#1}{\urlprefix }}%
\providecommand \urlprefix  [0]{URL }%
\providecommand \Eprint [0]{\href }%
\providecommand \doibase [0]{http://dx.doi.org/}%
\providecommand \selectlanguage [0]{\@gobble}%
\providecommand \bibinfo  [0]{\@secondoftwo}%
\providecommand \bibfield  [0]{\@secondoftwo}%
\providecommand \translation [1]{[#1]}%
\providecommand \BibitemOpen [0]{}%
\providecommand \bibitemStop [0]{}%
\providecommand \bibitemNoStop [0]{.\EOS\space}%
\providecommand \EOS [0]{\spacefactor3000\relax}%
\providecommand \BibitemShut  [1]{\csname bibitem#1\endcsname}%
\let\auto@bib@innerbib\@empty
%</preamble>
\bibitem [{\citenamefont {Knappe}\ \emph {et~al.}(2004)\citenamefont {Knappe},
  \citenamefont {Shah}, \citenamefont {Schwindt}, \citenamefont {Hollberg},
  \citenamefont {Kitching}, \citenamefont {Liew},\ and\ \citenamefont
  {Moreland}}]{Knappe2004}%
  \BibitemOpen
  \bibfield  {author} {\bibinfo {author} {\bibfnamefont {S.}~\bibnamefont
  {Knappe}}, \bibinfo {author} {\bibfnamefont {V.}~\bibnamefont {Shah}},
  \bibinfo {author} {\bibfnamefont {P.~D.~D.}\ \bibnamefont {Schwindt}},
  \bibinfo {author} {\bibfnamefont {L.}~\bibnamefont {Hollberg}}, \bibinfo
  {author} {\bibfnamefont {J.}~\bibnamefont {Kitching}}, \bibinfo {author}
  {\bibfnamefont {L.-A.}\ \bibnamefont {Liew}}, \ and\ \bibinfo {author}
  {\bibfnamefont {J.}~\bibnamefont {Moreland}},\ }\bibfield  {title} {\enquote
  {\bibinfo {title} {A microfabricated atomic clock},}\ }\href {\doibase
  10.1063/1.1787942} {\bibfield  {journal} {\bibinfo  {journal} {Applied
  Physics Letters}\ }\textbf {\bibinfo {volume} {85}},\ \bibinfo {pages}
  {1460--1462} (\bibinfo {year} {2004})}\BibitemShut {NoStop}%
\bibitem [{\citenamefont {Kitching}(2018)}]{Kitching2018}%
  \BibitemOpen
  \bibfield  {author} {\bibinfo {author} {\bibfnamefont {J.}~\bibnamefont
  {Kitching}},\ }\bibfield  {title} {\enquote {\bibinfo {title} {Chip-scale
  atomic devices},}\ }\href {\doibase 10.1063/1.5026238} {\bibfield  {journal}
  {\bibinfo  {journal} {Applied Physics Reviews}\ }\textbf {\bibinfo {volume}
  {5}},\ \bibinfo {pages} {031302} (\bibinfo {year} {2018})}\BibitemShut
  {NoStop}%
\bibitem [{\citenamefont {Carl\'e}\ \emph {et~al.}(2021)\citenamefont
  {Carl\'e}, \citenamefont {Petersen}, \citenamefont {Passilly}, \citenamefont
  {Hafiz}, \citenamefont {de~Clercq},\ and\ \citenamefont
  {Boudot}}]{Carle2021_microfab_clock}%
  \BibitemOpen
  \bibfield  {author} {\bibinfo {author} {\bibfnamefont {C.}~\bibnamefont
  {Carl\'e}}, \bibinfo {author} {\bibfnamefont {M.}~\bibnamefont {Petersen}},
  \bibinfo {author} {\bibfnamefont {N.}~\bibnamefont {Passilly}}, \bibinfo
  {author} {\bibfnamefont {M.~A.}\ \bibnamefont {Hafiz}}, \bibinfo {author}
  {\bibfnamefont {E.}~\bibnamefont {de~Clercq}}, \ and\ \bibinfo {author}
  {\bibfnamefont {R.}~\bibnamefont {Boudot}},\ }\bibfield  {title} {\enquote
  {\bibinfo {title} {Exploring the use of {R}amsey {CPT} spectroscopy for a
  microcell-based atomic clock},}\ }\href {\doibase 10.1109/tuffc.2021.3085249}
  {\bibfield  {journal} {\bibinfo  {journal} {{IEEE} Transactions on
  Ultrasonics, Ferroelectrics, and Frequency Control}\ ,\ \bibinfo {pages}
  {1--1}} (\bibinfo {year} {2021})}\BibitemShut {NoStop}%
\bibitem [{\citenamefont {Kozlova}, \citenamefont {Gu\'erandel},\ and\
  \citenamefont {de~Clercq}(2011)}]{Kozlova2011_bufer_gas}%
  \BibitemOpen
  \bibfield  {author} {\bibinfo {author} {\bibfnamefont {O.}~\bibnamefont
  {Kozlova}}, \bibinfo {author} {\bibfnamefont {S.}~\bibnamefont
  {Gu\'erandel}}, \ and\ \bibinfo {author} {\bibfnamefont {E.}~\bibnamefont
  {de~Clercq}},\ }\bibfield  {title} {\enquote {\bibinfo {title} {Temperature
  and pressure shift of the cs clock transition in the presence of buffer
  gases: Ne, $\mathrm{N}_{2}$, $\mathrm{Ar}$},}\ }\href {\doibase
  10.1103/PhysRevA.83.062714} {\bibfield  {journal} {\bibinfo  {journal}
  {Physical Review A}\ }\textbf {\bibinfo {volume} {83}},\ \bibinfo {pages}
  {062714} (\bibinfo {year} {2011})}\BibitemShut {NoStop}%
\bibitem [{\citenamefont {Hasegawa}\ \emph {et~al.}(2011)\citenamefont
  {Hasegawa}, \citenamefont {Chutani}, \citenamefont {Gorecki}, \citenamefont
  {Boudot}, \citenamefont {Dziuban}, \citenamefont {Giordano}, \citenamefont
  {Clatot},\ and\ \citenamefont {Mauri}}]{Hasegawa2011_buffer_gas}%
  \BibitemOpen
  \bibfield  {author} {\bibinfo {author} {\bibfnamefont {M.}~\bibnamefont
  {Hasegawa}}, \bibinfo {author} {\bibfnamefont {R.}~\bibnamefont {Chutani}},
  \bibinfo {author} {\bibfnamefont {C.}~\bibnamefont {Gorecki}}, \bibinfo
  {author} {\bibfnamefont {R.}~\bibnamefont {Boudot}}, \bibinfo {author}
  {\bibfnamefont {P.}~\bibnamefont {Dziuban}}, \bibinfo {author} {\bibfnamefont
  {V.}~\bibnamefont {Giordano}}, \bibinfo {author} {\bibfnamefont
  {S.}~\bibnamefont {Clatot}}, \ and\ \bibinfo {author} {\bibfnamefont
  {L.}~\bibnamefont {Mauri}},\ }\bibfield  {title} {\enquote {\bibinfo {title}
  {Microfabrication of cesium vapor cells with buffer gas for {MEMS} atomic
  clocks},}\ }\href {\doibase 10.1016/j.sna.2011.02.039} {\bibfield  {journal}
  {\bibinfo  {journal} {Sensors and Actuators A: Physical}\ }\textbf {\bibinfo
  {volume} {167}},\ \bibinfo {pages} {594--601} (\bibinfo {year}
  {2011})}\BibitemShut {NoStop}%
\bibitem [{\citenamefont {Seltzer}\ and\ \citenamefont
  {Romalis}(2009)}]{Seltzer2009_cell_coating}%
  \BibitemOpen
  \bibfield  {author} {\bibinfo {author} {\bibfnamefont {S.~J.}\ \bibnamefont
  {Seltzer}}\ and\ \bibinfo {author} {\bibfnamefont {M.~V.}\ \bibnamefont
  {Romalis}},\ }\bibfield  {title} {\enquote {\bibinfo {title}
  {High-temperature alkali vapor cells with antirelaxation surface coatings},}\
  }\href {\doibase 10.1063/1.3236649} {\bibfield  {journal} {\bibinfo
  {journal} {Journal of Applied Physics}\ }\textbf {\bibinfo {volume} {106}},\
  \bibinfo {pages} {114905} (\bibinfo {year} {2009})}\BibitemShut {NoStop}%
\bibitem [{\citenamefont {Eckel}\ \emph {et~al.}(2018)\citenamefont {Eckel},
  \citenamefont {Barker}, \citenamefont {Fedchak}, \citenamefont {Klimov},
  \citenamefont {Norrgard}, \citenamefont {Scherschligt}, \citenamefont
  {Makrides},\ and\ \citenamefont {Tiesinga}}]{Eckel_2018}%
  \BibitemOpen
  \bibfield  {author} {\bibinfo {author} {\bibfnamefont {S.}~\bibnamefont
  {Eckel}}, \bibinfo {author} {\bibfnamefont {D.~S.}\ \bibnamefont {Barker}},
  \bibinfo {author} {\bibfnamefont {J.~A.}\ \bibnamefont {Fedchak}}, \bibinfo
  {author} {\bibfnamefont {N.~N.}\ \bibnamefont {Klimov}}, \bibinfo {author}
  {\bibfnamefont {E.}~\bibnamefont {Norrgard}}, \bibinfo {author}
  {\bibfnamefont {J.}~\bibnamefont {Scherschligt}}, \bibinfo {author}
  {\bibfnamefont {C.}~\bibnamefont {Makrides}}, \ and\ \bibinfo {author}
  {\bibfnamefont {E.}~\bibnamefont {Tiesinga}},\ }\bibfield  {title} {\enquote
  {\bibinfo {title} {Challenges to miniaturizing cold atom technology for
  deployable vacuum metrology},}\ }\href {\doibase 10.1088/1681-7575/aadbe4}
  {\bibfield  {journal} {\bibinfo  {journal} {Metrologia}\ }\textbf {\bibinfo
  {volume} {55}},\ \bibinfo {pages} {S182--S193} (\bibinfo {year}
  {2018})}\BibitemShut {NoStop}%
\bibitem [{\citenamefont {Rushton}, \citenamefont {Aldous},\ and\ \citenamefont
  {Himsworth}(2014)}]{Rushton2014}%
  \BibitemOpen
  \bibfield  {author} {\bibinfo {author} {\bibfnamefont {J.~A.}\ \bibnamefont
  {Rushton}}, \bibinfo {author} {\bibfnamefont {M.}~\bibnamefont {Aldous}}, \
  and\ \bibinfo {author} {\bibfnamefont {M.~D.}\ \bibnamefont {Himsworth}},\
  }\bibfield  {title} {\enquote {\bibinfo {title} {Contributed review: The
  feasibility of a fully miniaturized magneto-optical trap for portable
  ultracold quantum technology},}\ }\href {\doibase 10.1063/1.4904066}
  {\bibfield  {journal} {\bibinfo  {journal} {Review of Scientific
  Instruments}\ }\textbf {\bibinfo {volume} {85}},\ \bibinfo {pages} {121501}
  (\bibinfo {year} {2014})}\BibitemShut {NoStop}%
\bibitem [{\citenamefont {Marlow}\ and\ \citenamefont
  {Scherer}(2021)}]{marlow2021review}%
  \BibitemOpen
  \bibfield  {author} {\bibinfo {author} {\bibfnamefont {B.~L.~S.}\
  \bibnamefont {Marlow}}\ and\ \bibinfo {author} {\bibfnamefont {D.~R.}\
  \bibnamefont {Scherer}},\ }\bibfield  {title} {\enquote {\bibinfo {title} {A
  review of commercial and emerging atomic frequency standards},}\ }\href
  {\doibase 10.1109/tuffc.2021.3049713} {\bibfield  {journal} {\bibinfo
  {journal} {{IEEE} Transactions on Ultrasonics, Ferroelectrics, and Frequency
  Control}\ }\textbf {\bibinfo {volume} {68}},\ \bibinfo {pages} {2007--2022}
  (\bibinfo {year} {2021})}\BibitemShut {NoStop}%
\bibitem [{\citenamefont {Muquans}()}]{MuClock}%
  \BibitemOpen
  \bibfield  {author} {\bibinfo {author} {\bibnamefont {Muquans}},\ }\href
  {https://www.muquans.com/product/muclock/} {\enquote {\bibinfo {title}
  {Muclock data sheet},}\ }\bibinfo {howpublished} {Available
  Online}\BibitemShut {NoStop}%
\bibitem [{\citenamefont {Burrow}\ \emph {et~al.}(2021)\citenamefont {Burrow},
  \citenamefont {Osborn}, \citenamefont {Boughton}, \citenamefont {Mirando},
  \citenamefont {Burt}, \citenamefont {Griffin}, \citenamefont {Arnold},\ and\
  \citenamefont {Riis}}]{burrow2021}%
  \BibitemOpen
  \bibfield  {author} {\bibinfo {author} {\bibfnamefont {O.~S.}\ \bibnamefont
  {Burrow}}, \bibinfo {author} {\bibfnamefont {P.~F.}\ \bibnamefont {Osborn}},
  \bibinfo {author} {\bibfnamefont {E.}~\bibnamefont {Boughton}}, \bibinfo
  {author} {\bibfnamefont {F.}~\bibnamefont {Mirando}}, \bibinfo {author}
  {\bibfnamefont {D.~P.}\ \bibnamefont {Burt}}, \bibinfo {author}
  {\bibfnamefont {P.~F.}\ \bibnamefont {Griffin}}, \bibinfo {author}
  {\bibfnamefont {A.~S.}\ \bibnamefont {Arnold}}, \ and\ \bibinfo {author}
  {\bibfnamefont {E.}~\bibnamefont {Riis}},\ }\href@noop {} {\enquote {\bibinfo
  {title} {Centilitre-scale vacuum chamber for compact ultracold quantum
  technologies},}\ } (\bibinfo {year} {2021}),\ \Eprint
  {http://arxiv.org/abs/2101.07851} {arXiv:2101.07851} \BibitemShut {NoStop}%
\bibitem [{\citenamefont {Little}\ \emph {et~al.}(2021)\citenamefont {Little},
  \citenamefont {Hoth}, \citenamefont {Christensen}, \citenamefont {Walker},
  \citenamefont {Smet}, \citenamefont {Biedermann}, \citenamefont {Lee},\ and\
  \citenamefont {Schwindt}}]{little2021}%
  \BibitemOpen
  \bibfield  {author} {\bibinfo {author} {\bibfnamefont {B.~J.}\ \bibnamefont
  {Little}}, \bibinfo {author} {\bibfnamefont {G.~W.}\ \bibnamefont {Hoth}},
  \bibinfo {author} {\bibfnamefont {J.}~\bibnamefont {Christensen}}, \bibinfo
  {author} {\bibfnamefont {C.}~\bibnamefont {Walker}}, \bibinfo {author}
  {\bibfnamefont {D.~J.~D.}\ \bibnamefont {Smet}}, \bibinfo {author}
  {\bibfnamefont {G.~W.}\ \bibnamefont {Biedermann}}, \bibinfo {author}
  {\bibfnamefont {J.}~\bibnamefont {Lee}}, \ and\ \bibinfo {author}
  {\bibfnamefont {P.~D.~D.}\ \bibnamefont {Schwindt}},\ }\href@noop {}
  {\enquote {\bibinfo {title} {A passively pumped vacuum package sustaining
  cold atoms for more than 200 days},}\ } (\bibinfo {year} {2021}),\ \Eprint
  {http://arxiv.org/abs/2101.01051} {arXiv:2101.01051} \BibitemShut {NoStop}%
\bibitem [{\citenamefont {Nichols}\ \emph {et~al.}(2020)\citenamefont
  {Nichols}, \citenamefont {Nofs}, \citenamefont {Viray}, \citenamefont {Ma},
  \citenamefont {Paradis},\ and\ \citenamefont {Raithel}}]{ball_lens_MOT}%
  \BibitemOpen
  \bibfield  {author} {\bibinfo {author} {\bibfnamefont {C.~S.}\ \bibnamefont
  {Nichols}}, \bibinfo {author} {\bibfnamefont {L.~M.}\ \bibnamefont {Nofs}},
  \bibinfo {author} {\bibfnamefont {M.~A.}\ \bibnamefont {Viray}}, \bibinfo
  {author} {\bibfnamefont {L.}~\bibnamefont {Ma}}, \bibinfo {author}
  {\bibfnamefont {E.}~\bibnamefont {Paradis}}, \ and\ \bibinfo {author}
  {\bibfnamefont {G.}~\bibnamefont {Raithel}},\ }\bibfield  {title} {\enquote
  {\bibinfo {title} {Magneto-optical trap with millimeter ball lenses},}\
  }\href {\doibase 10.1103/PhysRevApplied.14.044013} {\bibfield  {journal}
  {\bibinfo  {journal} {Physical Review Applied}\ }\textbf {\bibinfo {volume}
  {14}},\ \bibinfo {pages} {044013} (\bibinfo {year} {2020})}\BibitemShut
  {NoStop}%
\bibitem [{\citenamefont {Nshii}\ \emph {et~al.}(2013)\citenamefont {Nshii},
  \citenamefont {Vangeleyn}, \citenamefont {Cotter}, \citenamefont {Griffin},
  \citenamefont {Hinds}, \citenamefont {Ironside}, \citenamefont {See},
  \citenamefont {Sinclair}, \citenamefont {Riis},\ and\ \citenamefont
  {Arnold}}]{Nshii2013}%
  \BibitemOpen
  \bibfield  {author} {\bibinfo {author} {\bibfnamefont {C.~C.}\ \bibnamefont
  {Nshii}}, \bibinfo {author} {\bibfnamefont {M.}~\bibnamefont {Vangeleyn}},
  \bibinfo {author} {\bibfnamefont {J.~P.}\ \bibnamefont {Cotter}}, \bibinfo
  {author} {\bibfnamefont {P.~F.}\ \bibnamefont {Griffin}}, \bibinfo {author}
  {\bibfnamefont {E.~A.}\ \bibnamefont {Hinds}}, \bibinfo {author}
  {\bibfnamefont {C.~N.}\ \bibnamefont {Ironside}}, \bibinfo {author}
  {\bibfnamefont {P.}~\bibnamefont {See}}, \bibinfo {author} {\bibfnamefont
  {A.~G.}\ \bibnamefont {Sinclair}}, \bibinfo {author} {\bibfnamefont
  {E.}~\bibnamefont {Riis}}, \ and\ \bibinfo {author} {\bibfnamefont {A.~S.}\
  \bibnamefont {Arnold}},\ }\bibfield  {title} {\enquote {\bibinfo {title} {A
  surface-patterned chip as a strong source of ultracold atoms for quantum
  technologies},}\ }\href {\doibase 10.1038/nnano.2013.47} {\bibfield
  {journal} {\bibinfo  {journal} {Nature Nanotechnology}\ }\textbf {\bibinfo
  {volume} {8}},\ \bibinfo {pages} {321--324} (\bibinfo {year}
  {2013})}\BibitemShut {NoStop}%
\bibitem [{\citenamefont {Ravenhall}, \citenamefont {Yuen},\ and\ \citenamefont
  {Foot}(2021)}]{Ravenhall21}%
  \BibitemOpen
  \bibfield  {author} {\bibinfo {author} {\bibfnamefont {S.}~\bibnamefont
  {Ravenhall}}, \bibinfo {author} {\bibfnamefont {B.}~\bibnamefont {Yuen}}, \
  and\ \bibinfo {author} {\bibfnamefont {C.}~\bibnamefont {Foot}},\ }\bibfield
  {title} {\enquote {\bibinfo {title} {High-flux, adjustable, compact cold-atom
  source},}\ }\href {\doibase 10.1364/OE.423662} {\bibfield  {journal}
  {\bibinfo  {journal} {Optics Express}\ }\textbf {\bibinfo {volume} {29}},\
  \bibinfo {pages} {21143--21159} (\bibinfo {year} {2021})}\BibitemShut
  {NoStop}%
\bibitem [{\citenamefont {Kang}\ \emph {et~al.}(2019)\citenamefont {Kang},
  \citenamefont {Moore}, \citenamefont {McGilligan}, \citenamefont {Mott},
  \citenamefont {Mis}, \citenamefont {Roper}, \citenamefont {Donley},\ and\
  \citenamefont {Kitching}}]{Kang:19}%
  \BibitemOpen
  \bibfield  {author} {\bibinfo {author} {\bibfnamefont {S.}~\bibnamefont
  {Kang}}, \bibinfo {author} {\bibfnamefont {K.~R.}\ \bibnamefont {Moore}},
  \bibinfo {author} {\bibfnamefont {J.~P.}\ \bibnamefont {McGilligan}},
  \bibinfo {author} {\bibfnamefont {R.}~\bibnamefont {Mott}}, \bibinfo {author}
  {\bibfnamefont {A.}~\bibnamefont {Mis}}, \bibinfo {author} {\bibfnamefont
  {C.}~\bibnamefont {Roper}}, \bibinfo {author} {\bibfnamefont {E.~A.}\
  \bibnamefont {Donley}}, \ and\ \bibinfo {author} {\bibfnamefont
  {J.}~\bibnamefont {Kitching}},\ }\bibfield  {title} {\enquote {\bibinfo
  {title} {Magneto-optic trap using a reversible, solid-state alkali-metal
  source},}\ }\href {\doibase 10.1364/OL.44.003002} {\bibfield  {journal}
  {\bibinfo  {journal} {Optics Letters}\ }\textbf {\bibinfo {volume} {44}},\
  \bibinfo {pages} {3002--3005} (\bibinfo {year} {2019})}\BibitemShut {NoStop}%
\bibitem [{\citenamefont {McGehee}\ \emph {et~al.}(2021)\citenamefont
  {McGehee}, \citenamefont {Zhu}, \citenamefont {Barker}, \citenamefont
  {Westly}, \citenamefont {Yulaev}, \citenamefont {Klimov}, \citenamefont
  {Agrawal}, \citenamefont {Eckel}, \citenamefont {Aksyuk},\ and\ \citenamefont
  {McClelland}}]{McGehee_2021}%
  \BibitemOpen
  \bibfield  {author} {\bibinfo {author} {\bibfnamefont {W.~R.}\ \bibnamefont
  {McGehee}}, \bibinfo {author} {\bibfnamefont {W.}~\bibnamefont {Zhu}},
  \bibinfo {author} {\bibfnamefont {D.~S.}\ \bibnamefont {Barker}}, \bibinfo
  {author} {\bibfnamefont {D.}~\bibnamefont {Westly}}, \bibinfo {author}
  {\bibfnamefont {A.}~\bibnamefont {Yulaev}}, \bibinfo {author} {\bibfnamefont
  {N.}~\bibnamefont {Klimov}}, \bibinfo {author} {\bibfnamefont
  {A.}~\bibnamefont {Agrawal}}, \bibinfo {author} {\bibfnamefont
  {S.}~\bibnamefont {Eckel}}, \bibinfo {author} {\bibfnamefont
  {V.}~\bibnamefont {Aksyuk}}, \ and\ \bibinfo {author} {\bibfnamefont {J.~J.}\
  \bibnamefont {McClelland}},\ }\bibfield  {title} {\enquote {\bibinfo {title}
  {Magneto-optical trapping using planar optics},}\ }\href {\doibase
  10.1088/1367-2630/abdce3} {\bibfield  {journal} {\bibinfo  {journal} {New
  Journal of Physics}\ }\textbf {\bibinfo {volume} {23}},\ \bibinfo {pages}
  {013021} (\bibinfo {year} {2021})}\BibitemShut {NoStop}%
\bibitem [{\citenamefont {Sitaram}\ \emph
  {et~al.}(2020{\natexlab{a}})\citenamefont {Sitaram}, \citenamefont {Elgee},
  \citenamefont {Campbell}, \citenamefont {Klimov}, \citenamefont {Eckel},\
  and\ \citenamefont {Barker}}]{eckelsr}%
  \BibitemOpen
  \bibfield  {author} {\bibinfo {author} {\bibfnamefont {A.}~\bibnamefont
  {Sitaram}}, \bibinfo {author} {\bibfnamefont {P.~K.}\ \bibnamefont {Elgee}},
  \bibinfo {author} {\bibfnamefont {G.~K.}\ \bibnamefont {Campbell}}, \bibinfo
  {author} {\bibfnamefont {N.~N.}\ \bibnamefont {Klimov}}, \bibinfo {author}
  {\bibfnamefont {S.}~\bibnamefont {Eckel}}, \ and\ \bibinfo {author}
  {\bibfnamefont {D.~S.}\ \bibnamefont {Barker}},\ }\bibfield  {title}
  {\enquote {\bibinfo {title} {Confinement of an alkaline-earth element in a
  grating magneto-optical trap},}\ }\href {\doibase 10.1063/5.0019551}
  {\bibfield  {journal} {\bibinfo  {journal} {Review of Scientific
  Instruments}\ }\textbf {\bibinfo {volume} {91}},\ \bibinfo {pages} {103202}
  (\bibinfo {year} {2020}{\natexlab{a}})}\BibitemShut {NoStop}%
\bibitem [{\citenamefont {McGilligan}\ \emph {et~al.}(2020)\citenamefont
  {McGilligan}, \citenamefont {Moore}, \citenamefont {Dellis}, \citenamefont
  {Martinez}, \citenamefont {de~Clercq}, \citenamefont {Griffin}, \citenamefont
  {Arnold}, \citenamefont {Riis}, \citenamefont {Boudot},\ and\ \citenamefont
  {Kitching}}]{McGilligan2020}%
  \BibitemOpen
  \bibfield  {author} {\bibinfo {author} {\bibfnamefont {J.~P.}\ \bibnamefont
  {McGilligan}}, \bibinfo {author} {\bibfnamefont {K.~R.}\ \bibnamefont
  {Moore}}, \bibinfo {author} {\bibfnamefont {A.}~\bibnamefont {Dellis}},
  \bibinfo {author} {\bibfnamefont {G.~D.}\ \bibnamefont {Martinez}}, \bibinfo
  {author} {\bibfnamefont {E.}~\bibnamefont {de~Clercq}}, \bibinfo {author}
  {\bibfnamefont {P.~F.}\ \bibnamefont {Griffin}}, \bibinfo {author}
  {\bibfnamefont {A.~S.}\ \bibnamefont {Arnold}}, \bibinfo {author}
  {\bibfnamefont {E.}~\bibnamefont {Riis}}, \bibinfo {author} {\bibfnamefont
  {R.}~\bibnamefont {Boudot}}, \ and\ \bibinfo {author} {\bibfnamefont
  {J.}~\bibnamefont {Kitching}},\ }\bibfield  {title} {\enquote {\bibinfo
  {title} {Laser cooling in a chip-scale platform},}\ }\href {\doibase
  10.1063/5.0014658} {\bibfield  {journal} {\bibinfo  {journal} {Applied
  Physics Letters}\ }\textbf {\bibinfo {volume} {117}},\ \bibinfo {pages}
  {054001} (\bibinfo {year} {2020})}\BibitemShut {NoStop}%
\bibitem [{\citenamefont {Sheludko}\ \emph {et~al.}(2008)\citenamefont
  {Sheludko}, \citenamefont {Bell}, \citenamefont {Anderson}, \citenamefont
  {Hofmann}, \citenamefont {Vredenbregt},\ and\ \citenamefont
  {Scholten}}]{scholten}%
  \BibitemOpen
  \bibfield  {author} {\bibinfo {author} {\bibfnamefont {D.~V.}\ \bibnamefont
  {Sheludko}}, \bibinfo {author} {\bibfnamefont {S.~C.}\ \bibnamefont {Bell}},
  \bibinfo {author} {\bibfnamefont {R.}~\bibnamefont {Anderson}}, \bibinfo
  {author} {\bibfnamefont {C.~S.}\ \bibnamefont {Hofmann}}, \bibinfo {author}
  {\bibfnamefont {E.~J.~D.}\ \bibnamefont {Vredenbregt}}, \ and\ \bibinfo
  {author} {\bibfnamefont {R.~E.}\ \bibnamefont {Scholten}},\ }\bibfield
  {title} {\enquote {\bibinfo {title} {State-selective imaging of cold
  atoms},}\ }\href {\doibase 10.1103/PhysRevA.77.033401} {\bibfield  {journal}
  {\bibinfo  {journal} {Physical Review A}\ }\textbf {\bibinfo {volume} {77}},\
  \bibinfo {pages} {033401} (\bibinfo {year} {2008})}\BibitemShut {NoStop}%
\bibitem [{\citenamefont {Imhof}\ \emph {et~al.}(2017)\citenamefont {Imhof},
  \citenamefont {Stuhl}, \citenamefont {Kasch}, \citenamefont {Kroese},
  \citenamefont {Olson},\ and\ \citenamefont {Squires}}]{imhof}%
  \BibitemOpen
  \bibfield  {author} {\bibinfo {author} {\bibfnamefont {E.}~\bibnamefont
  {Imhof}}, \bibinfo {author} {\bibfnamefont {B.~K.}\ \bibnamefont {Stuhl}},
  \bibinfo {author} {\bibfnamefont {B.}~\bibnamefont {Kasch}}, \bibinfo
  {author} {\bibfnamefont {B.}~\bibnamefont {Kroese}}, \bibinfo {author}
  {\bibfnamefont {S.~E.}\ \bibnamefont {Olson}}, \ and\ \bibinfo {author}
  {\bibfnamefont {M.~B.}\ \bibnamefont {Squires}},\ }\bibfield  {title}
  {\enquote {\bibinfo {title} {Two-dimensional grating magneto-optical trap},}\
  }\href {\doibase 10.1103/PhysRevA.96.033636} {\bibfield  {journal} {\bibinfo
  {journal} {Phys. Rev. A}\ }\textbf {\bibinfo {volume} {96}},\ \bibinfo
  {pages} {033636} (\bibinfo {year} {2017})}\BibitemShut {NoStop}%
\bibitem [{\citenamefont {Franssen}\ \emph {et~al.}(2019)\citenamefont
  {Franssen}, \citenamefont {de~Raadt}, \citenamefont {van Ninhuijs},\ and\
  \citenamefont {Luiten}}]{jimion}%
  \BibitemOpen
  \bibfield  {author} {\bibinfo {author} {\bibfnamefont {J.~G.~H.}\
  \bibnamefont {Franssen}}, \bibinfo {author} {\bibfnamefont {T.~C.~H.}\
  \bibnamefont {de~Raadt}}, \bibinfo {author} {\bibfnamefont {M.~A.~W.}\
  \bibnamefont {van Ninhuijs}}, \ and\ \bibinfo {author} {\bibfnamefont
  {O.~J.}\ \bibnamefont {Luiten}},\ }\bibfield  {title} {\enquote {\bibinfo
  {title} {Compact ultracold electron source based on a grating magneto-optical
  trap},}\ }\href {\doibase 10.1103/PhysRevAccelBeams.22.023401} {\bibfield
  {journal} {\bibinfo  {journal} {Phys. Rev. Accel. Beams}\ }\textbf {\bibinfo
  {volume} {22}},\ \bibinfo {pages} {023401} (\bibinfo {year}
  {2019})}\BibitemShut {NoStop}%
\bibitem [{\citenamefont {Barker}\ \emph {et~al.}(2019)\citenamefont {Barker},
  \citenamefont {Norrgard}, \citenamefont {Klimov}, \citenamefont {Fedchak},
  \citenamefont {Scherschligt},\ and\ \citenamefont {Eckel}}]{barkerNIST}%
  \BibitemOpen
  \bibfield  {author} {\bibinfo {author} {\bibfnamefont {D.}~\bibnamefont
  {Barker}}, \bibinfo {author} {\bibfnamefont {E.}~\bibnamefont {Norrgard}},
  \bibinfo {author} {\bibfnamefont {N.}~\bibnamefont {Klimov}}, \bibinfo
  {author} {\bibfnamefont {J.}~\bibnamefont {Fedchak}}, \bibinfo {author}
  {\bibfnamefont {J.}~\bibnamefont {Scherschligt}}, \ and\ \bibinfo {author}
  {\bibfnamefont {S.}~\bibnamefont {Eckel}},\ }\bibfield  {title} {\enquote
  {\bibinfo {title} {Single-beam zeeman slower and magneto-optical trap using a
  nanofabricated grating},}\ }\href {\doibase 10.1103/PhysRevApplied.11.064023}
  {\bibfield  {journal} {\bibinfo  {journal} {Phys. Rev. Applied}\ }\textbf
  {\bibinfo {volume} {11}},\ \bibinfo {pages} {064023} (\bibinfo {year}
  {2019})}\BibitemShut {NoStop}%
\bibitem [{\citenamefont {Sitaram}\ \emph
  {et~al.}(2020{\natexlab{b}})\citenamefont {Sitaram}, \citenamefont {Elgee},
  \citenamefont {Campbell}, \citenamefont {Klimov}, \citenamefont {Eckel},\
  and\ \citenamefont {Barker}}]{srgmot}%
  \BibitemOpen
  \bibfield  {author} {\bibinfo {author} {\bibfnamefont {A.}~\bibnamefont
  {Sitaram}}, \bibinfo {author} {\bibfnamefont {P.~K.}\ \bibnamefont {Elgee}},
  \bibinfo {author} {\bibfnamefont {G.~K.}\ \bibnamefont {Campbell}}, \bibinfo
  {author} {\bibfnamefont {N.~N.}\ \bibnamefont {Klimov}}, \bibinfo {author}
  {\bibfnamefont {S.}~\bibnamefont {Eckel}}, \ and\ \bibinfo {author}
  {\bibfnamefont {D.~S.}\ \bibnamefont {Barker}},\ }\bibfield  {title}
  {\enquote {\bibinfo {title} {Confinement of an alkaline-earth element in a
  grating magneto-optical trap},}\ }\href {\doibase 10.1063/5.0019551}
  {\bibfield  {journal} {\bibinfo  {journal} {Review of Scientific
  Instruments}\ }\textbf {\bibinfo {volume} {91}},\ \bibinfo {pages} {103202}
  (\bibinfo {year} {2020}{\natexlab{b}})},\ \Eprint
  {http://arxiv.org/abs/https://doi.org/10.1063/5.0019551}
  {https://doi.org/10.1063/5.0019551} \BibitemShut {NoStop}%
\bibitem [{\citenamefont {Dellis}\ \emph {et~al.}(2016)\citenamefont {Dellis},
  \citenamefont {Shah}, \citenamefont {Donley}, \citenamefont {Knappe},\ and\
  \citenamefont {Kitching}}]{dellis2016lowHe}%
  \BibitemOpen
  \bibfield  {author} {\bibinfo {author} {\bibfnamefont {A.~T.}\ \bibnamefont
  {Dellis}}, \bibinfo {author} {\bibfnamefont {V.}~\bibnamefont {Shah}},
  \bibinfo {author} {\bibfnamefont {E.~A.}\ \bibnamefont {Donley}}, \bibinfo
  {author} {\bibfnamefont {S.}~\bibnamefont {Knappe}}, \ and\ \bibinfo {author}
  {\bibfnamefont {J.}~\bibnamefont {Kitching}},\ }\bibfield  {title} {\enquote
  {\bibinfo {title} {Low helium permeation cells for atomic microsystems
  technology},}\ }\href {\doibase 10.1364/ol.41.002775} {\bibfield  {journal}
  {\bibinfo  {journal} {Optics Letters}\ }\textbf {\bibinfo {volume} {41}},\
  \bibinfo {pages} {2775} (\bibinfo {year} {2016})}\BibitemShut {NoStop}%
\bibitem [{\citenamefont {Boudot}\ \emph {et~al.}(2020)\citenamefont {Boudot},
  \citenamefont {McGilligan}, \citenamefont {Moore}, \citenamefont {Maurice},
  \citenamefont {Martinez}, \citenamefont {Hansen}, \citenamefont {de~Clercq},\
  and\ \citenamefont {Kitching}}]{BoudotPassive}%
  \BibitemOpen
  \bibfield  {author} {\bibinfo {author} {\bibfnamefont {R.}~\bibnamefont
  {Boudot}}, \bibinfo {author} {\bibfnamefont {J.~P.}\ \bibnamefont
  {McGilligan}}, \bibinfo {author} {\bibfnamefont {K.~R.}\ \bibnamefont
  {Moore}}, \bibinfo {author} {\bibfnamefont {V.}~\bibnamefont {Maurice}},
  \bibinfo {author} {\bibfnamefont {G.~D.}\ \bibnamefont {Martinez}}, \bibinfo
  {author} {\bibfnamefont {A.}~\bibnamefont {Hansen}}, \bibinfo {author}
  {\bibfnamefont {E.}~\bibnamefont {de~Clercq}}, \ and\ \bibinfo {author}
  {\bibfnamefont {J.}~\bibnamefont {Kitching}},\ }\bibfield  {title} {\enquote
  {\bibinfo {title} {Enhanced observation time of magneto-optical traps using
  micro-machined non-evaporable getter pumps},}\ }\href
  {https://doi.org/10.1038/s41598-020-73605-z} {\bibfield  {journal} {\bibinfo
  {journal} {Scientific Reports}\ }\textbf {\bibinfo {volume} {10}},\ \bibinfo
  {pages} {16590} (\bibinfo {year} {2020})}\BibitemShut {NoStop}%
\bibitem [{\citenamefont {Arpornthip}, \citenamefont {Sackett},\ and\
  \citenamefont {Hughes}(2012)}]{arpornthip2012pressuremeasurment}%
  \BibitemOpen
  \bibfield  {author} {\bibinfo {author} {\bibfnamefont {T.}~\bibnamefont
  {Arpornthip}}, \bibinfo {author} {\bibfnamefont {C.~A.}\ \bibnamefont
  {Sackett}}, \ and\ \bibinfo {author} {\bibfnamefont {K.~J.}\ \bibnamefont
  {Hughes}},\ }\bibfield  {title} {\enquote {\bibinfo {title} {Vacuum-pressure
  measurement using a magneto-optical trap},}\ }\href {\doibase
  10.1103/PhysRevA.85.033420} {\bibfield  {journal} {\bibinfo  {journal}
  {Physical Review A}\ }\textbf {\bibinfo {volume} {85}},\ \bibinfo {pages}
  {033420} (\bibinfo {year} {2012})}\BibitemShut {NoStop}%
\bibitem [{\citenamefont {Moore}\ \emph {et~al.}(2015)\citenamefont {Moore},
  \citenamefont {Lee}, \citenamefont {Findlay}, \citenamefont {Torralbo-Campo},
  \citenamefont {Bruce},\ and\ \citenamefont
  {Cassettari}}]{moore2015pressuremeasurement}%
  \BibitemOpen
  \bibfield  {author} {\bibinfo {author} {\bibfnamefont {R.~W.~G.}\
  \bibnamefont {Moore}}, \bibinfo {author} {\bibfnamefont {L.~A.}\ \bibnamefont
  {Lee}}, \bibinfo {author} {\bibfnamefont {E.~A.}\ \bibnamefont {Findlay}},
  \bibinfo {author} {\bibfnamefont {L.}~\bibnamefont {Torralbo-Campo}},
  \bibinfo {author} {\bibfnamefont {G.~D.}\ \bibnamefont {Bruce}}, \ and\
  \bibinfo {author} {\bibfnamefont {D.}~\bibnamefont {Cassettari}},\ }\bibfield
   {title} {\enquote {\bibinfo {title} {Measurement of vacuum pressure with a
  magneto-optical trap: A pressure-rise method},}\ }\href {\doibase
  10.1063/1.4928154} {\bibfield  {journal} {\bibinfo  {journal} {Review of
  Scientific Instruments}\ }\textbf {\bibinfo {volume} {86}},\ \bibinfo {pages}
  {093108} (\bibinfo {year} {2015})}\BibitemShut {NoStop}%
\bibitem [{\citenamefont {McGilligan}\ \emph {et~al.}(2017)\citenamefont
  {McGilligan}, \citenamefont {Griffin}, \citenamefont {Elvin}, \citenamefont
  {Ingleby}, \citenamefont {Riis},\ and\ \citenamefont
  {Arnold}}]{McGilligan2017}%
  \BibitemOpen
  \bibfield  {author} {\bibinfo {author} {\bibfnamefont {J.~P.}\ \bibnamefont
  {McGilligan}}, \bibinfo {author} {\bibfnamefont {P.~F.}\ \bibnamefont
  {Griffin}}, \bibinfo {author} {\bibfnamefont {R.}~\bibnamefont {Elvin}},
  \bibinfo {author} {\bibfnamefont {S.~J.}\ \bibnamefont {Ingleby}}, \bibinfo
  {author} {\bibfnamefont {E.}~\bibnamefont {Riis}}, \ and\ \bibinfo {author}
  {\bibfnamefont {A.~S.}\ \bibnamefont {Arnold}},\ }\bibfield  {title}
  {\enquote {\bibinfo {title} {Grating chips for quantum technologies},}\
  }\href {\doibase 10.1038/s41598-017-00254-0} {\bibfield  {journal} {\bibinfo
  {journal} {Scientific Reports}\ }\textbf {\bibinfo {volume} {7}},\ \bibinfo
  {pages} {384} (\bibinfo {year} {2017})}\BibitemShut {NoStop}%
\bibitem [{\citenamefont {Smith}\ \emph {et~al.}(2011)\citenamefont {Smith},
  \citenamefont {Aigner}, \citenamefont {Hofferberth}, \citenamefont {Gring},
  \citenamefont {Andersson}, \citenamefont {Wildermuth}, \citenamefont
  {Kr\"{u}ger}, \citenamefont {Schneider}, \citenamefont {Schumm},\ and\
  \citenamefont {Schmiedmayer}}]{smithabsorption}%
  \BibitemOpen
  \bibfield  {author} {\bibinfo {author} {\bibfnamefont {D.~A.}\ \bibnamefont
  {Smith}}, \bibinfo {author} {\bibfnamefont {S.}~\bibnamefont {Aigner}},
  \bibinfo {author} {\bibfnamefont {S.}~\bibnamefont {Hofferberth}}, \bibinfo
  {author} {\bibfnamefont {M.}~\bibnamefont {Gring}}, \bibinfo {author}
  {\bibfnamefont {M.}~\bibnamefont {Andersson}}, \bibinfo {author}
  {\bibfnamefont {S.}~\bibnamefont {Wildermuth}}, \bibinfo {author}
  {\bibfnamefont {P.}~\bibnamefont {Kr\"{u}ger}}, \bibinfo {author}
  {\bibfnamefont {S.}~\bibnamefont {Schneider}}, \bibinfo {author}
  {\bibfnamefont {T.}~\bibnamefont {Schumm}}, \ and\ \bibinfo {author}
  {\bibfnamefont {J.}~\bibnamefont {Schmiedmayer}},\ }\bibfield  {title}
  {\enquote {\bibinfo {title} {Absorption imaging of ultracold atoms on atom
  chips},}\ }\href {\doibase 10.1364/OE.19.008471} {\bibfield  {journal}
  {\bibinfo  {journal} {Opt. Express}\ }\textbf {\bibinfo {volume} {19}},\
  \bibinfo {pages} {8471--8485} (\bibinfo {year} {2011})}\BibitemShut {NoStop}%
\bibitem [{\citenamefont {McGilligan}\ \emph {et~al.}(2015)\citenamefont
  {McGilligan}, \citenamefont {Griffin}, \citenamefont {Riis},\ and\
  \citenamefont {Arnold}}]{mcgilligan15}%
  \BibitemOpen
  \bibfield  {author} {\bibinfo {author} {\bibfnamefont {J.~P.}\ \bibnamefont
  {McGilligan}}, \bibinfo {author} {\bibfnamefont {P.~F.}\ \bibnamefont
  {Griffin}}, \bibinfo {author} {\bibfnamefont {E.}~\bibnamefont {Riis}}, \
  and\ \bibinfo {author} {\bibfnamefont {A.~S.}\ \bibnamefont {Arnold}},\
  }\bibfield  {title} {\enquote {\bibinfo {title} {Phase-space properties of
  magneto-optical traps utilising micro-fabricated gratings.}}\ }\href
  {\doibase 10.1364/OE.23.008948} {\bibfield  {journal} {\bibinfo  {journal}
  {Optics Express}\ }\textbf {\bibinfo {volume} {23}},\ \bibinfo {pages}
  {8948--8959} (\bibinfo {year} {2015})}\BibitemShut {NoStop}%
\bibitem [{\citenamefont {Lindquist}, \citenamefont {Stephens},\ and\
  \citenamefont {Wieman}(1992)}]{wieman1992}%
  \BibitemOpen
  \bibfield  {author} {\bibinfo {author} {\bibfnamefont {K.}~\bibnamefont
  {Lindquist}}, \bibinfo {author} {\bibfnamefont {M.}~\bibnamefont {Stephens}},
  \ and\ \bibinfo {author} {\bibfnamefont {C.}~\bibnamefont {Wieman}},\
  }\bibfield  {title} {\enquote {\bibinfo {title} {Experimental and theoretical
  study of the vapor-cell zeeman optical trap},}\ }\href {\doibase
  10.1103/PhysRevA.46.4082} {\bibfield  {journal} {\bibinfo  {journal}
  {Physical Review A}\ }\textbf {\bibinfo {volume} {46}},\ \bibinfo {pages}
  {4082--4090} (\bibinfo {year} {1992})}\BibitemShut {NoStop}%
\bibitem [{\citenamefont {Gibble}, \citenamefont {Kasapi},\ and\ \citenamefont
  {Chu}(1992)}]{Gibble:92}%
  \BibitemOpen
  \bibfield  {author} {\bibinfo {author} {\bibfnamefont {K.~E.}\ \bibnamefont
  {Gibble}}, \bibinfo {author} {\bibfnamefont {S.}~\bibnamefont {Kasapi}}, \
  and\ \bibinfo {author} {\bibfnamefont {S.}~\bibnamefont {Chu}},\ }\bibfield
  {title} {\enquote {\bibinfo {title} {Improved magneto-optic trapping in a
  vapor cell},}\ }\href {\doibase 10.1364/OL.17.000526} {\bibfield  {journal}
  {\bibinfo  {journal} {Optics letters}\ }\textbf {\bibinfo {volume} {17}},\
  \bibinfo {pages} {526--528} (\bibinfo {year} {1992})}\BibitemShut {NoStop}%
\bibitem [{\citenamefont {Hoth}, \citenamefont {Donley},\ and\ \citenamefont
  {Kitching}(2013)}]{Hoth13}%
  \BibitemOpen
  \bibfield  {author} {\bibinfo {author} {\bibfnamefont {G.~W.}\ \bibnamefont
  {Hoth}}, \bibinfo {author} {\bibfnamefont {E.~A.}\ \bibnamefont {Donley}}, \
  and\ \bibinfo {author} {\bibfnamefont {J.}~\bibnamefont {Kitching}},\
  }\bibfield  {title} {\enquote {\bibinfo {title} {Atom number in magneto-optic
  traps with millimeter scale laser beams},}\ }\href {\doibase
  10.1364/ol.38.000661} {\bibfield  {journal} {\bibinfo  {journal} {Optics
  Letters}\ }\textbf {\bibinfo {volume} {38}},\ \bibinfo {pages} {661}
  (\bibinfo {year} {2013})}\BibitemShut {NoStop}%
\bibitem [{\citenamefont {Pollock}\ \emph {et~al.}(2011)\citenamefont
  {Pollock}, \citenamefont {Cotter}, \citenamefont {Laliotis}, \citenamefont
  {Ramirez-Martinez},\ and\ \citenamefont {Hinds}}]{Pollock_2011}%
  \BibitemOpen
  \bibfield  {author} {\bibinfo {author} {\bibfnamefont {S.}~\bibnamefont
  {Pollock}}, \bibinfo {author} {\bibfnamefont {J.~P.}\ \bibnamefont {Cotter}},
  \bibinfo {author} {\bibfnamefont {A.}~\bibnamefont {Laliotis}}, \bibinfo
  {author} {\bibfnamefont {F.}~\bibnamefont {Ramirez-Martinez}}, \ and\
  \bibinfo {author} {\bibfnamefont {E.~A.}\ \bibnamefont {Hinds}},\ }\bibfield
  {title} {\enquote {\bibinfo {title} {Characteristics of integrated
  magneto-optical traps for atom chips},}\ }\href {\doibase
  10.1088/1367-2630/13/4/043029} {\bibfield  {journal} {\bibinfo  {journal}
  {New Journal of Physics}\ }\textbf {\bibinfo {volume} {13}},\ \bibinfo
  {pages} {043029} (\bibinfo {year} {2011})}\BibitemShut {NoStop}%
\bibitem [{\citenamefont {Niu}\ \emph {et~al.}(2018)\citenamefont {Niu},
  \citenamefont {Guo}, \citenamefont {Zhan}, \citenamefont {Chen},
  \citenamefont {Liu},\ and\ \citenamefont {Zhou}}]{fringe_removal_Niu2018}%
  \BibitemOpen
  \bibfield  {author} {\bibinfo {author} {\bibfnamefont {L.}~\bibnamefont
  {Niu}}, \bibinfo {author} {\bibfnamefont {X.}~\bibnamefont {Guo}}, \bibinfo
  {author} {\bibfnamefont {Y.}~\bibnamefont {Zhan}}, \bibinfo {author}
  {\bibfnamefont {X.}~\bibnamefont {Chen}}, \bibinfo {author} {\bibfnamefont
  {W.~M.}\ \bibnamefont {Liu}}, \ and\ \bibinfo {author} {\bibfnamefont
  {X.}~\bibnamefont {Zhou}},\ }\bibfield  {title} {\enquote {\bibinfo {title}
  {Optimized fringe removal algorithm for absorption images},}\ }\href
  {\doibase 10.1063/1.5040669} {\bibfield  {journal} {\bibinfo  {journal}
  {Applied Physics Letters}\ }\textbf {\bibinfo {volume} {113}},\ \bibinfo
  {pages} {144103} (\bibinfo {year} {2018})}\BibitemShut {NoStop}%
\bibitem [{\citenamefont {Xiong}, \citenamefont {Long},\ and\ \citenamefont
  {Parker}(2020)}]{fringe_removal_Xiong2020}%
  \BibitemOpen
  \bibfield  {author} {\bibinfo {author} {\bibfnamefont {F.}~\bibnamefont
  {Xiong}}, \bibinfo {author} {\bibfnamefont {Y.}~\bibnamefont {Long}}, \ and\
  \bibinfo {author} {\bibfnamefont {C.~V.}\ \bibnamefont {Parker}},\ }\bibfield
   {title} {\enquote {\bibinfo {title} {Enhanced principle component method for
  fringe removal in cold atom images},}\ }\href {\doibase 10.1364/josab.391297}
  {\bibfield  {journal} {\bibinfo  {journal} {Journal of the Optical Society of
  America B}\ }\textbf {\bibinfo {volume} {37}},\ \bibinfo {pages} {2041}
  (\bibinfo {year} {2020})}\BibitemShut {NoStop}%
\end{thebibliography}%

\end{document}